\documentclass[iicol,sn-basic]{sn-jnl}

\usepackage{natbib,multirow,xspace}
\usepackage{xcolor}

\usepackage{graphicx}%
\usepackage{multirow}%
\usepackage{amsmath,amssymb,amsfonts}%
\usepackage{amsthm}%
\usepackage{mathrsfs}%
\usepackage[title]{appendix}%
\usepackage{xcolor}%
\usepackage{textcomp}%
\usepackage{manyfoot}%
\usepackage{booktabs}%
\usepackage{algorithm}%
\usepackage{algorithmicx}%
\usepackage{algpseudocode}%
\usepackage{listings}%
\usepackage{lineno}

\def\apjl{{ApJL}}
\def\apjs{{ApJS}}
\def\ao{{Appl.~Opt.}}

\def\aap{{A\&A}}

\newcommand{\nustar}{\emph{NuSTAR}\xspace}
\newcommand{\swift}{\emph{Swift}\xspace}
\newcommand{\suzaku}{\emph{Suzaku}\xspace}
\newcommand{\chandra}{\emph{Chandra}\xspace}
\newcommand{\nicer}{\emph{NICER}\xspace}

\newcommand{\rxte}{\emph{RXTE}\xspace}

\newcommand{\xmm}{\emph{XMM-Newton}\xspace}
\newcommand{\xrism}{\emph{XRISM}\xspace}
\newcommand{\athena}{\emph{Athena}\xspace}

\newcommand{\ms}{$M_{\odot}$\xspace}
\newcommand{\mdot}{$\dot{\mathrm{m}}$\xspace}

\newcommand{\lumcgs}{ergs~s$^{-1}$\xspace}
\newcommand{\rin}{$R_{\rm in}$\xspace}
\newcommand{\rg}{$R_{g}$\xspace}
\newcommand{\risco}{$R_{\mathrm{ISCO}}$\xspace}

\newcommand{\relxill}{{\sc relxill}\xspace}
\newcommand{\relxillns}{{\sc relxillNS}\xspace}

\newcommand{\xillverco}{{\sc xillverCO}\xspace}

\begin{document}

\title[Article Title]{Reflecting on Accretion in  Neutron Star Low-Mass X-ray Binaries}



\author[1]{\fnm{Renee M.} \sur{Ludlam}}

{ \affil[1]{\orgdiv{Department of Physics and Astronomy}, \orgname{Wayne State University}, \orgaddress{\street{666 W. Hancock St.}, \city{Detroit}, \postcode{48201}, \state{MI}, \country{U.S.A}}}}

\abstract{Neutron star low-mass X-ray binaries accrete via Roche-lobe overflow from a stellar companion that is $\lesssim$\,1 M$_{\odot}$. The accretion disk in these systems can be externally illuminated by X-rays that are  reprocessed by the accreting material into an emergent reflection spectrum comprised of emission lines superimposed onto the reprocessed continuum. Due to proximity to the compact object, strong gravity effects are imparted to the reflection spectrum that can be modeled to infer properties of the NS itself and other aspects of the accreting system.
This short review discusses the field of reflection modeling in neutron star low-mass X-ray binary systems with the intention to highlight the work that was awarded the 2023 AAS Newton Lacy Pierce Prize, but also to consolidate key information as a reference for those entering this subfield.}


\maketitle

\section{Background} \label{sec:intro}
Accretion is a ubiquitous process in the universe occurring on both small scales and large scales (e.g., planet formation to supermassive black holes at the center of galaxies). Generally, this process yields the formation of a flat disk structure as a consequence of angular momentum conservation (i.e., an accretion disk).
When an accretion disk is externally illuminated by X-rays, these photons are absorbed and reprocessed by the material in the disk prior to being re-emitted. The emergent photons form a reprocessed continuum spectrum with narrow emission and absorption lines superimposed with characteristics indicative of the physical properties of the material in the disk (e.g., the ionization state, density, and composition, etc). This reprocessed emission emerging from the accretion disk is referred to in the literature as the `reflection spectrum' \citep{Ross2005,Garcia2010}.

\begin{figure}
\centering
\includegraphics[width=0.48\textwidth,trim=5 10 0 0,clip]{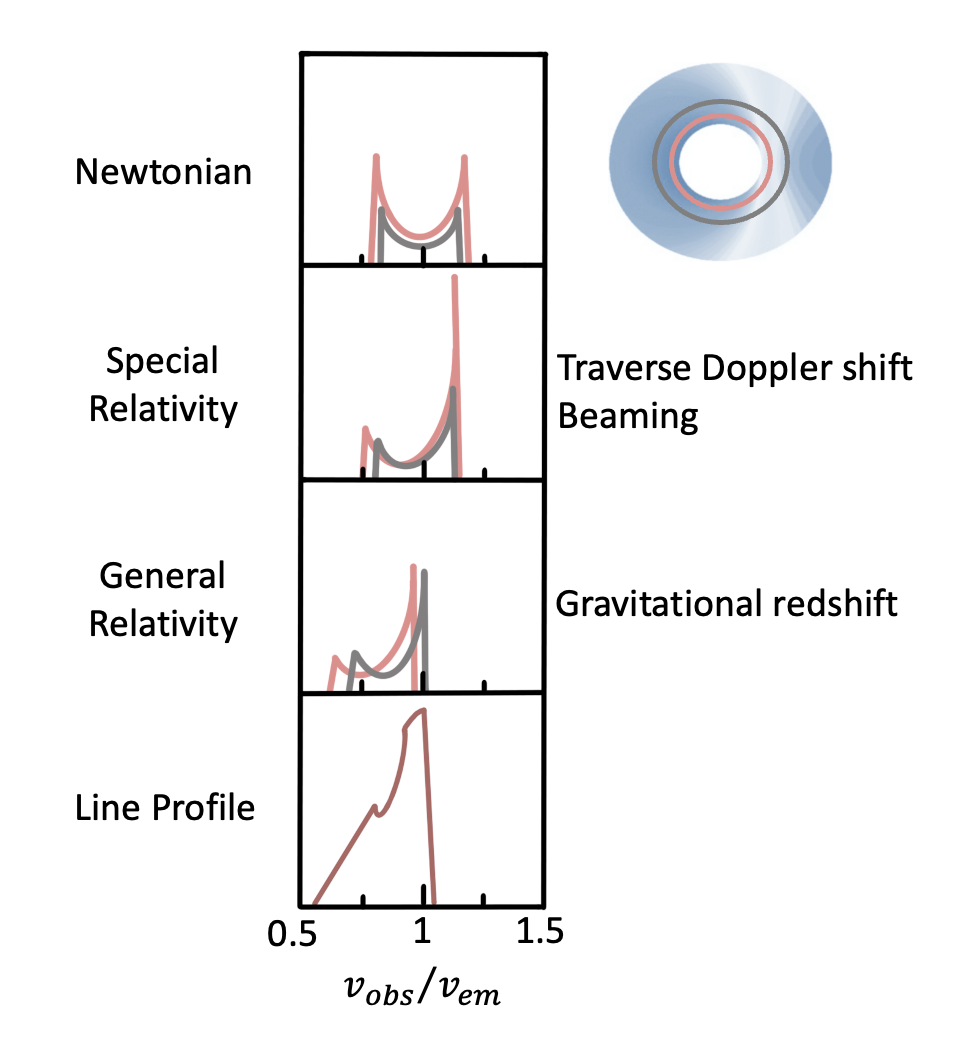}
\caption{The impact of different effects on the intrinsically narrow Fe~K line profile based on the location within the disk where the emission arises. The annulus in the top right represents the accretion disk with two circle to denote the emission radii shown in the panels (pink close to the compact object and grey further out in the disk). The x-axis shows the broadening effects in terms of the observed line frequency over the emitted frequency (equivalent to energy observed with respect to the energy emitted). Figure adapted from \cite{Fabian00}. Classical Newtonian motion leads to Doppler shifts that create a two-horned profile. The rapid rate at which the matter is rotating within the disk leads to Special relativistic effects that creates the asymmetric line profile as photons are beamed into and out of our line of sight. the large gravitational potential well that the disk exists in close to the compact object results in General relativistic effects of gravitational redshift dampening the emission and shifting the photons to lower frequency as they lose energy to escape the strong gravity. The overall line profile is the summation of the emission and effects from different radii.}
\label{fig:effects}
\end{figure}

The most prominent emission line in the reflection spectrum is from the iron (Fe) K$\alpha$ transition. This is due to the fact that Fe is one of the most tightly bound nuclei and able to keep its inner-shell electrons in these highly energetic environments. Other key features of the reflection spectrum are emission lines at lower energy (e.g., oxygen lyman alpha; \citealt{madej11, madej14, Ludlam19b, Ludlam21, moutard23}, or the Fe L-shell transition; \citealt{ludlam18}), the Fe K edge near 7\,keV, and the Compton hump near 20\,keV due to illuminating photons undergoing Compton scattering when interacting with material in the disk. The intrinsically narrow emission lines are broadened due to Doppler, special, and general relativistic effects. The strength of these effects depend on the location in which the line emission originates in the accretion disk \citep{Fabian00, Dauser10}. Figure~\ref{fig:effects} shows the impact of these effects based on location in the accretion disk. 

Relativistically broadened Fe lines indicative of reflection from the inner accretion disk were first observed in Active Galactic Nuclei (AGNs) \citep{tanaka95}, which are accreting supermassive black holes (BHs) at the center of galaxies. They were later identified in galactic BH X-ray binaries (XRBs) \citep{miller02} and more recently in neutron star (NS) XRBs \citep{Bhattacharyya07, cackett08}. This suggests a similar accretion geometry in these systems despite the mass difference of the compact accretors vary by orders of magnitude and the presence of a surface in the case of accreting NS. 
The degree of broadening measured from the Fe line profile in these systems indicated that the line emission indeed arose from the innermost region of the accretion disk.

\begin{figure}
\centering
\includegraphics[width=0.49\textwidth,trim=10 6 0 2,clip]{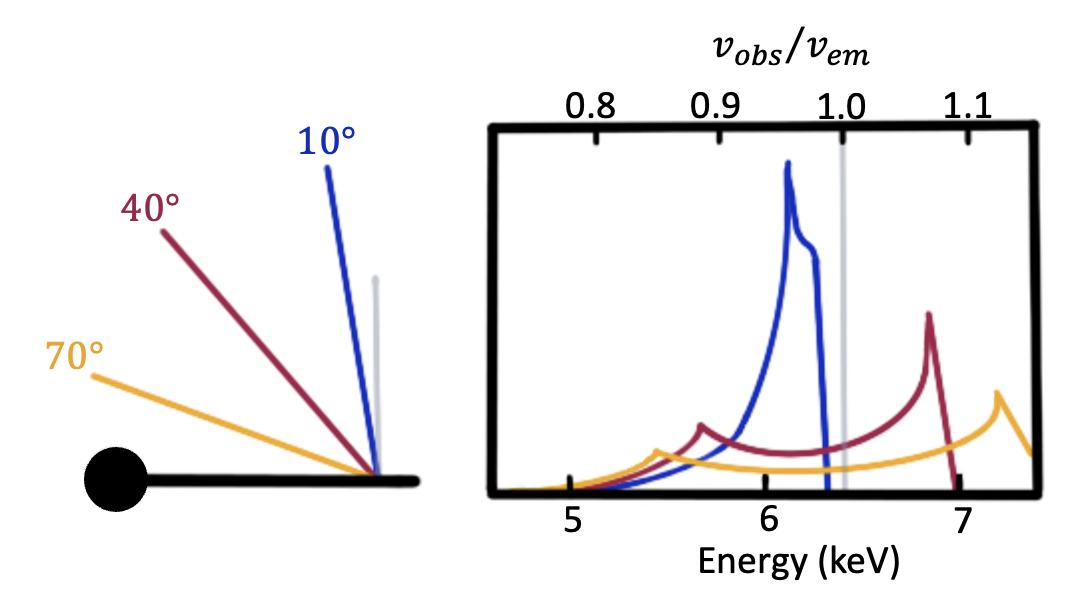}
\caption{The effect of inclination on the Fe~K line profile adapted from \cite{Dauser10}. Higher inclinations lead to a more extended blue wing of the Fe line profile. Though the example shows the effect of inclination for a maximally spinning BH at different viewing angles, the effect of inclination of the line profile is similar for lower spinning systems, however the gravitational redshift would be less pronounced.}
\label{fig:inc}
\end{figure}

In the strong gravity regime, general relativity sets the location of the last stable circular orbit of a test particle orbiting an object \citep{bardeen72}. For a static (non-rotating) compact object, the location of the innermost stable circular orbit (\risco) is 6 gravitational radii ($R_g=GM/c^2$). In the case where a central compact object is rotating prograde (i.e., orbits in the disk are in the same direction as the compact objects angular momentum vector), this distorts the fabric of space-time and supports orbits closer to the BH or NS. Since the gravitational redshift effects on the red-wing of the line profile become stronger closer to the compact object \citep{Fabian00}, modeling of the Fe line profile enables a measure of the location of the inner disk radius. Additionally, the blue-wing of the line profile is impacted by Doppler effects that strengthen at higher inclinations \citep{Dauser10} (see Figure~\ref{fig:inc}), therefore reflection modeling also enables a measure of the inclination of the system.

Measuring the location of the inner disk in these systems has provided important inferences on the properties of these compact objects. In the case of BHs, this has allowed the measure of the dimensionless spin parameter ($a=cJ/GM^2$), which is one of two physical quantities needed to completely describe an astrophysical BH --- the other being mass  (see, e.g., \citealt{miller09, reynolds14, reynolds21} for review).
For NSs, the inner disk radius provides an upper limit on the NS radius \citep{miller13, Ludlam17a, Ludlam22} (which has important implications for the equation of state of ultra-dense, cold matter: \citealt{Lattimer01, Lattimer04, Lattimer2011}), magnetic field strength \citep{ibragimov2009, cackett09, papitto09, degenaar14, Degenaar16, king16, Ludlam17c}, and extent of the boundary layer region extending from the surface of the NS \citep{ibragimov2009, king16, Chiang16, Ludlam21}. 

\begin{figure*}
\centering
\includegraphics[width=\textwidth,trim=2 6 0 2,clip]{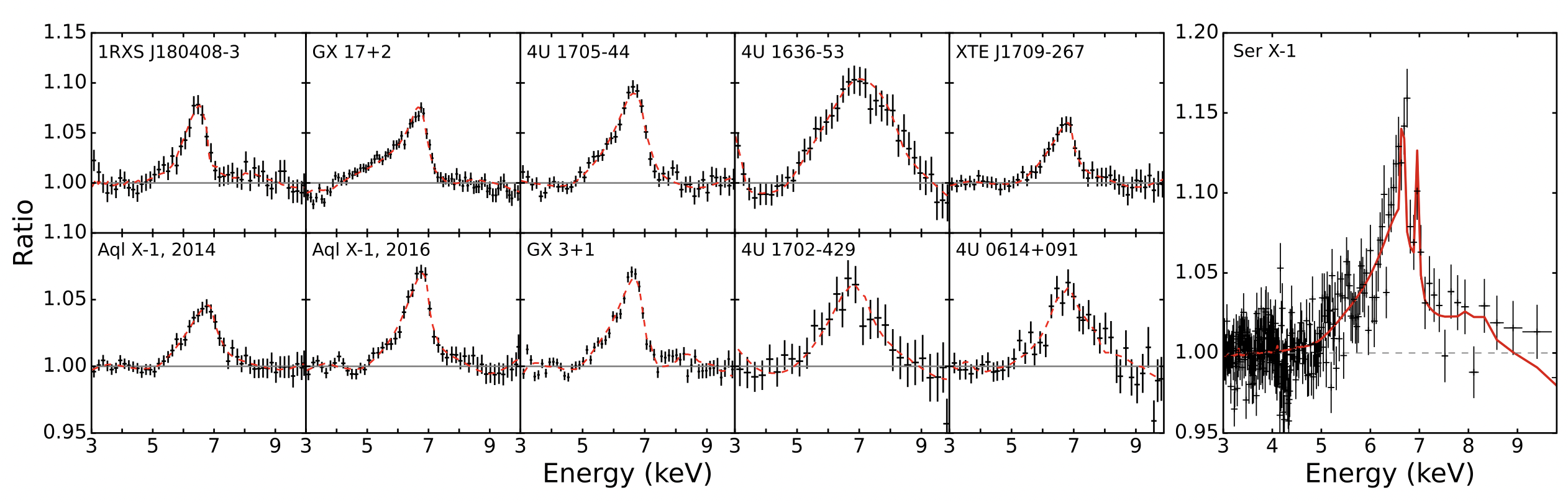}
\caption{({\it Left:}) Gallery of Fe line profiles observed with \nustar in accreting NS LMXBs. The ratio is the data to a simple continuum model. The red dashed line indicates the Fe line predicted from reflection modeling with the \relxill suite of models. The specifics for each source are given in their respective publication: 1RXS~J180408-3 \citep{Ludlam16}; GX~17+2, 4U~1705-44, and 4U~1636-53 \citep{Ludlam17a}; XTE~J1709-267 \citep{Ludlam17b}; Aquila X-1 (Aql~X-1) \citep{Ludlam17c}; GX 3+1, 4U~1702-429, and 4U~0614+091 \citep{Ludlam19a}. ({\it Right:}) The Fe line profile of Serpens X-1 (Ser X-1) observed with \nicer and reported in \cite{ludlam18}. The solid red line indicates the line profile predicted from fitting \relxillns to the data. The energy resolution of \nicer ($\sim137$\,eV at 6\,keV) provides additional detail regarding the structure of the Fe~line profile in this system. The double-peaked broad line is a result of high disk density ($\sim10^{19}$\,cm$^{-3}$) allowing both the Fe~{\sc XXV} and Fe~{\sc XXVI} K$_{\alpha}$ lines to fluoresce at a similar intensity. The low inclination of the system prevents the two lines from blurring together. }
\label{fig:FeLines}
\end{figure*}

\subsection{Controversy for Line Broadening}

When broadened Fe line profiles were discovered in BH and NS XRBs, there were alternative origins proposed for the broadening mechanism other than photon reprocessing from the inner disk region. One possible explanation was that the broadening was artificial and arose from pile-up effects in the CCD detectors used on the X-ray missions that were observing these bright galactic sources, e.g., \chandra \citep{chandra}, \swift \citep{Gehrels04}, \xmm \citep{xmm}, and \suzaku \citep{suzaku}. Pile-up occurs when two or more low-energy photons interact with the CCD during a single frame readout and register as one higher energy photon. Investigations \citep{yamada09, done10, ng10} showed that the line profile did change based upon how the data were extracted (e.g., excising the brightest regions of the detector), but this effect was ultimately demonstrated to artificially narrow the line emission \citep{miller10}, leading to underestimates of the inner disk radius location. Additionally, broadened Fe lines were detected with \rxte/PCA which has no pile-up effects \citep{garcia14b}, thus lending to the credibility of the broadening being astrophysical. Further advances in X-ray detectors, such as the CZT detectors on \nustar \citep{harrison13} and the silicon-drift detectors on \nicer \citep{Gendreau12}, eliminate pile-up effects (except at extremely high count rates of $10^{5}$\,counts\,s$^{-1}$ in the case of \nustar) and provide an unhindered view of line broadening effects. Figure~\ref{fig:FeLines} shows a gallery of Fe~line profiles observed with \nustar that are unbiased by pile-up effects, as well as a more recent example from \nicer.

Another proposed explanation for the broadening of the observed Fe line profiles was scattering interactions with winds as the photons left the system \citep{laurent07, titarchuk09, diaztrigo12}. However, these broadened lines were observed in hard spectral states for BH XRBs when winds were not present \citep{miller12} and no significant correlation was found between line width and column density of wind in different NS systems \citep{cackett13}. This supports a dynamical origin for the line broadening in these systems. Today the effects of the inner disk as the source of broadening in the reflection spectrum is a generally agreed upon idea.
However, there remains controversy over the proper spectral modeling description of the emission from different states, which is discussed further in the next section.

\subsection{Emission from NS LMXBs}


A NS low-mass X-ray binary (LMXB) accretes via Roche-lobe overflow from a stellar companion that is $\lesssim1$\,\ms (for a recent comprehensive reference on LMXBs, see \citealt{Bahramian23}). The accretion in these systems can be persistent or go through transient periods of accretion activity and quiescence (where little to no accretion occurs). The persistently accreting systems fall into one of two categories: `Z' or `atoll'. These names refer to characteristic shapes traced out when looking at the ratio of different energy bands (color-color diagrams) or ratio of the hard X-ray band to the soft X-ray band versus of the overall intensity (hardness-intensity diagram; \citealt{hasinger89}). The division between the two classes is thought to be driven in part by the overall level of mass accretion rate in the system since atolls accrete at a lower Eddington ratio than Z sources (${\sim} 0.01$--$0.5\,L_{\mathrm{Edd}}$ for atolls versus ${\sim} 0.5$--$1.0\,L_{\mathrm{Edd}}$ for Z sources: \citealt{vanderklis05}). Evidence supporting this comes from transiently accreting systems that switched between atoll-like behavior to Z-like behavior over the course of its outburst as the mass accretion rate changes (e.g., XTE\,J1701$-$462: \citealt{homan10}, and more recently 1A\,1744$-$361: \citealt{Ng2023}).

As mentioned previously, the presence of broadened Fe lines indicates a similar accretion geometry between BH and NS XRBs. BH XRBs exhibit a number of spectral states with varying accretion geometries as the overall accretion rate changes, which requires different models and parameter values in each state (see \citealt{done07} for a review). 
Unsurprisingly, NS LMXBs also exhibit a number of spectral states that vary considerably and require different models and spectral parameters to describe the continuum emission \citep{barret01} with the additional complexity of the presence of the NS surface that is not present in BH systems. Various models are used to describe the relative contribution from the presence of the corona \citep{Sunyaev1991}, accretion disk, surface of the NS and/or boundary layer region (where the material from the accretion disk meets the surface of the NS: \citealt{popham2001, inogamov99}) in each state. 

Transient NS LMXBs can exhibit very hard spectral states at the start of an outburst \citep{Ludlam16, parikh17, fiocchi19} where the spectrum is dominated by Comptonization from the corona region with little to no emission detected from the accretion disk or NS surface. This extreme hard state can be modeled with an absorbed power-law component with a photon index $\Gamma\sim1$ \citep{Ludlam16, parikh17} or two thermal Comptonization components assuming there is a two temperature coronal plasma with distinct populations of seed photons \citep{fiocchi19}. In the hard state, the spectrum is dominated by a hard Comptonization component from the corona with a steeper photon index of $\Gamma=1.5$--2.0 and the addition of a  soft thermal component with a temperature of $kT\lesssim1$\,keV \citep{barret00, church01}, which arises either from a single temperature blackbody component from the NS surface and/or boundary layer region, or from a multi-color disk blackbody emitted by the accretion disk.  

In the soft state, the spectrum becomes thermally dominated by the accretion disk emission or from emission near the NS with weakly Comptonized radiation. Model choices differ to describe the overall continuum in the soft state. Historically, this led to two classical descriptions in the literature: the ``Eastern'' and ``Western'' models. The Eastern model, after \cite{mitsuda89}, uses a multi-color disk blackbody for the accretion disk in combination with a Comptonized single-temperature blackbody component arising from Comptonized photons from a boundary layer region. The Western model, after \cite{white88},  uses a single-temperature blackbody to model emission from the NS or boundary layer with a Comptonized accretion disk component. 

A hybrid model was devised by \cite{Lin07} to build a coherent picture of the spectral evolution between the hard and soft state spectra using \textit{RXTE} observations of two transient atoll systems. In this model, the soft state assumes two thermal components (i.e., a single-temperature blackbody and an accretion disk blackbody) and weak Comptonization that is accounted for by a power-law component. The thermal components followed the expected $L_{x}\propto T^{4}$ relation as the mass accretion rate ($\dot{m} \propto L_{x}$) changed between states \citep{Lin07}. Although this hybrid double thermal model has been used in many NS LMXBs studies (e.g., \citealt{cackett08, cackett09, Lin10, miller13}), the various soft state model prescriptions are still utilized. A recent study used a model akin to the ``Eastern" model (i.e., thermal Comptonization from a blackbody and a multi-color accretion disk blackbody) for the X-ray spectra of the atoll 4U~1820-30 and found good agreement between the changes in the Comptonized blackbody with the observed jet variability in this system \citep{marino23}. Hence, spectral models that can describe data equally well in the X-ray band benefit from the additionally information provided by multi-wavelength data. In the absence of multi-wavelength data to support a choice of continuum model, it is difficult to ascertain which prescription of the spectra is appropriate and multiple model descriptions need to be explored. 

The exact geometry of the corona also remains a subject of mystery (see \citealt{Degenaar18} for some possible geometries). However, the recent operation of the {\it Imaging X-ray Polarization Explorer} ({\it IXPE}; \citealt{ixpe}) has enabled X-ray polarization measurements for a handful of accreting NSs which are beginning to shed light on the coronal orientation and presence of boundary layer regions in these systems (e.g., \citealt{Jayasurya23, Ursini23, Cocchi23, Farinelli23}). Similar to multi-wavelength studies, X-ray polarization measurements aid in building up our understanding of the accretion geometry when coupled with X-ray energy spectral studies. Although the choice of continuum impacts the interpretation of the accretion geometry and relative contribution of each component in the various spectral states, reflection studies have shown that the choice of continuum does not impact the results significantly, as long as the reprocessed emission is treated in a consistent manner with the assumed illumination component \citep{Coughenour18, Ludlam20}. 

\begin{figure*}
\centering
\includegraphics[width=\textwidth,trim=2 6 0 2,clip]{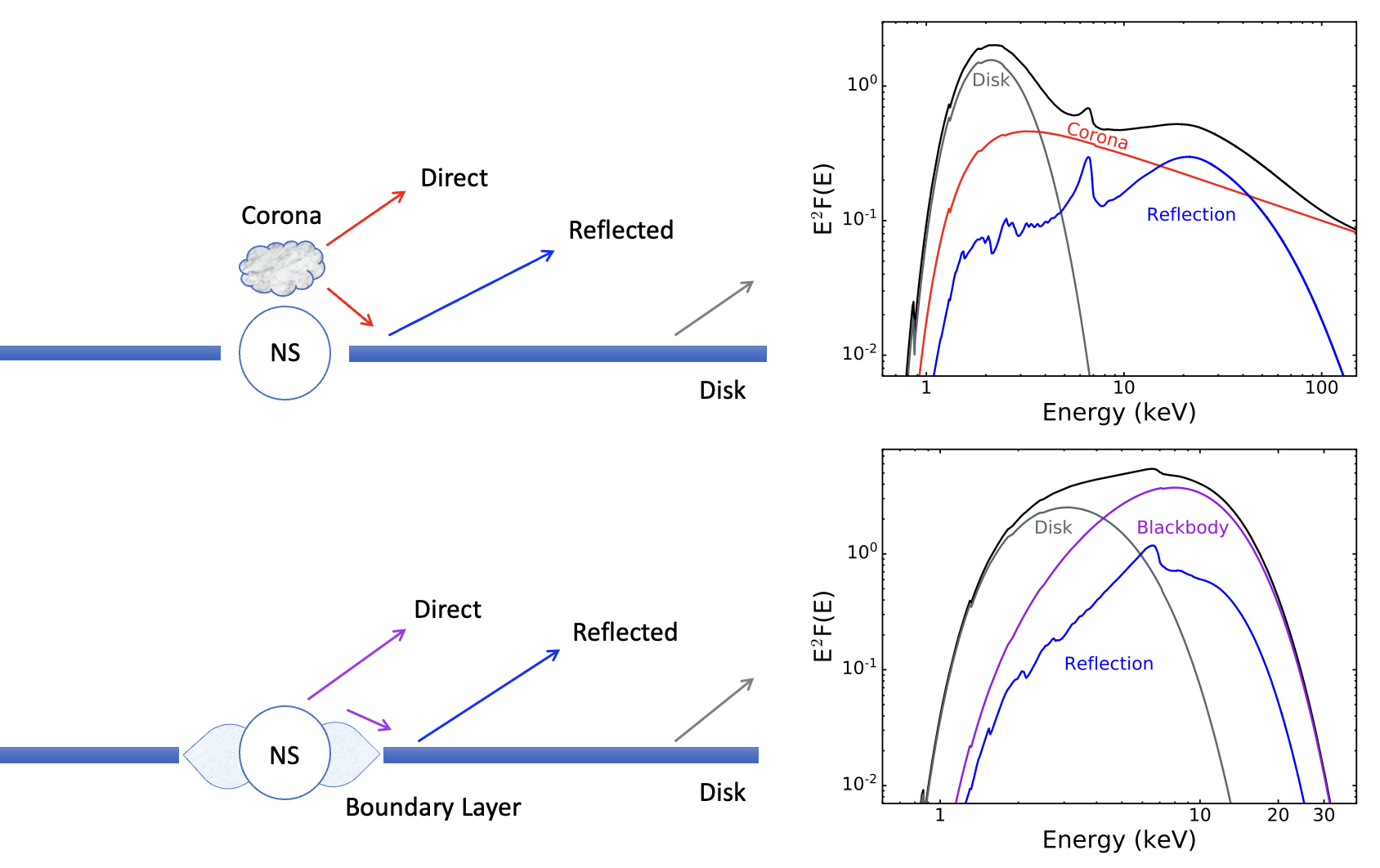}
\caption{Simplistic representation of emission of a NS accreting from a geometrically thin, optically thick accretion disk in a hard coronally-dominated spectral state (top) and soft thermally-dominated spectral state (bottom). The resulting unfolded model versus energy from these states are shown in the panels on the right using arbitrary values. The thermally dominated state has a significantly softer spectrum than a coronally-dominated state. This demonstrates how the external illumination component (either a power-law or blackbody irradiating the disk) shapes the reprocessed emission.}
\label{fig:toymodel}
\end{figure*}

\section{Reflection Modeling} \label{sec:model}

Modeling the reflection spectrum to infer properties of the accreting system originally used single line emission models \citep{Fabian1989, laor91}. However, we know that the reflection spectrum is a modified continuum with multiple atomic emission lines hence a full reflection model framework should be used to properly describe all the reprocessed emission. Eventually more sophisticated model grids were developed that encompassed the reprocessed continuum from irradiation of a constant density medium with varying illumination continua \citep{Magdziarz95, ross99, Garcia2010, garcia11, garcia13, Garcia22}.  Additionally, ad-hoc convolution reflection models were created that piece together expected shapes in different energy bands, such as {\sc rfxconv} \citep{done06, kolehmainen11}, which generates an angle-dependent reflection spectrum from a Comptonized input spectrum by combining the reflection emission from an ionized disk interpolated from {\sc reflionx} \citep{Ross2005} below 14\,keV (using the average 2–10\,keV power-law index) with Compton reflected emission from {\sc pexriv} \citep{Magdziarz95} above 14\,keV (using the average 12–14\,keV power-law index).

These reprocessed emission models needed to be blurred with convolution kernels to account for Doppler, special, and general relativistic effects, such as `rdblur' (modified from {\sc diskline} for non-spinning BHs: \citealt{Fabian1989}),  `kdblur' (modified from {\sc laor} for maximally rotating BHs: \citealt{laor91}), `kerrconv' \citep{Brenneman06}, and `relconv' \citep{Dauser10} (the latter two have the spin of the compact object is a free parameter). As the available models for the reflection spectrum and convolution kernel became sophisticated, the development of a self-consistent reflection model that traces the path of each photon from the system to construct the emergent spectrum was created in 2014 known as \relxill \citep{dauser14, garcia14}. The model convolves {\sc xillver} \citep{Garcia2010, garcia11, garcia13} reflection model tables with the blurring kernel of `relconv' \citep{Dauser10,Dauser2013}. Within \relxill, the model chooses the appropriate reflection spectrum at each disk radius for each emission angle calculated in a curved space-time, which consequently accurately captures the detailed dependence of the emitted reflection spectrum with the viewing angle. 

\subsection{Parameters of \relxill with respect to NS LMXBs}

A comprehensive explanation of parameter definitions in the \relxill suite of models is given on the developers website\footnote{\url{https://www.sternwarte.uni-erlangen.de/~dauser/research/relxill/relxill_docu_v1.0.pdf}}. 
We briefly cover these parameters here as similar parameters are used in other existing reflection models and how they relate to reflection studies in accreting NS systems.
The reflection grids are generated based on an illuminating continuum input (i.e., the `primary source') and thus result in different flavors of the model. For example, \relxill has a cut-off power-law with photon index $\Gamma$ and cut-off energy $E_{\rm cut}$ as the illuminating photon distribution, whereas \relxillns \citep{Garcia22} was designed assuming thermal illumination from the NS or boundary layer region with a temperature of $kT_{\rm bb}$. There is a flavor of \relxill with a Comptonization illuminating continuum known as {\sc relxillCp}, however this assumes the seed photons arise from the accretion disk with a fixed temperature of 0.01\,keV that is appropriate for BHs but too low for NS systems \citep{Ludlam19a}. Hence, {\sc rfxconv} is typically used when modeling reflection from Comptonization in NS systems.
Figure~\ref{fig:toymodel} provides a simplistic, toy-model representation of the accretion geometry in a hard and soft spectral state to demonstrate the difference in the overall resulting spectral shape given illumination by a power-law versus a blackbody component. \cite{Garcia22} provides a more in-depth comparison of the input models and resulting reflection spectral shape. 

The emissivity profile, $q$, in reflection modeling characterizes the illumination pattern of the accretion disk by the primary source and is a radially dependent prescription of $r^{-q}$. Whether the emissivity index is a positive or negative value depends on how the model was defined. Given that these models were designed for the extreme gravity near a BH, the models allow for two emissivity index profiles for illumination of the inner disk region where strong gravitational light bending effects cause steep profiles ($q_{1}$) and outer region of the accretion disk where effects are less pronounced ($q_{2}$) \citep{wilkins12}. The break radius ($\rm Rbr$) sets the location in the disk where the emissivity index changes from $q_1$ to $q_2$ in units of \rg. However, in NS systems, the gravitational effects are less extreme and a single emissivity profile is used (i.e., $q_1 = q_2$, hence $\rm Rbr$ is obsolete) with an expected value $q<5$ \citep{cackett10}. Indeed, theoretical emissivity profile calculations through ray-tracing simulations from a hotspot on the surface of the NS or boundary layer region support a single emissivity index close to the expected value for flat, Euclidean geometry \citep{wilkins18}.  Some studies elect to fix the emissivity index at $q=3$ for the flat, Euclidean geometry (e.g., \citealt{sleator16, Degenaar16, Ludlam17b, Ludlam17c, mondal18, mondal20, Coughenour18}), however, for the majority of studies where the emissivity index is left as a free parameter a value between $1.5<q<4$ is typically found (e.g., \citealt{miller13, matranga17, Ludlam16, Ludlam17a, ludlam18, Ludlam19a, Ludlam20, Ludlam21, ludlam23, saavedra23, moutard23}).

The extent of the accretion disk is set by the inner disk radius (\rin, in units of \risco or \rg depending on the model) and outer disk radius ($R_{\rm out}$, in units of \rg). \rin is typically left free to vary, whereas $R_{\rm out}$ is fixed at or near the maximum value allowed by the model. The inclination ($i$) of the system with respect to our line of sight is given in degrees. The cosmological redshift to the source is defined by $z$. However, for most of the analyses discussed here this parameter is fixed at 0 since the sources of interest are galactic. The normalization of the \relxill suite of models is discussed in detail in \citep{dauser16}.

As previously mentioned in \S\ref{sec:intro}, the dimensionless spin parameter ($a$) sets the location of the innermost stable circular orbit of the accretion disk (\risco). The majority of galactic NSs in LMXBs have spin $a \lesssim 0.3$ \citep{galloway2008, miller2011}. The difference in the location of \risco between $a=0$ and $a=0.3$ is $\sim1$\,\rg in the Kerr metric \citep{bardeen72}. Deviations from the Kerr metric may occur as rotation of the NS increases and induces a quadrupole moment as the NS becomes oblate, however, the Kerr metric remains a good approximation. The exact deviation from the Kerr metric depends upon the equation of state (EoS) which remains elusive, but at $a=0.3$ the deviations from an induced quadrupole moment is at most 10\% \citep{sibgatullin98}. This becomes a larger deviation ($\sim 25$\%, see Figure~1 in \citealt{sibgatullin98}) closer to rotation limit of the NS ($a\sim0.7$), where the centrifugal force overcomes the self-gravity of the NS. There is an approximation for spin relating to  measured spin periods of NSs in units of milliseconds ($a=0.47/P_{\rm ms}$: \citealt{braje00}), but it is important to note that this assumes a softish EOS of the ``FPS" model \citep{cook94} and canonical values for NS mass and radius ($M_{\rm NS}=1.4$\,\ms, $R_{\rm NS}=10$\,km). Given limitations with data quality and the degenerate nature between $a$ and \risco, the spin is fixed at a value in reflection modeling while the inner disk radius (\rin) is free to vary. Future mission concepts may provide the necessary data quality to determine both parameters \citep{ludlam23}.

These models allow for variation in the abundance of iron, $A_{\rm Fe}$, with respect to solar abundance. The value of $A_{\rm Fe}$ impacts the shape of the reprocessed continuum as well as the relative strength of the emission from different atomic species (e.g., see Figure~4 of \citealt{Garcia22}). A number of galactic accreting XRBs were inferred to have super-solar Fe abundance ($A_{\rm Fe}\gtrsim5$; e.g., \citealt{tomsick18, ludlam18, Ludlam19a, Ludlam17a, Connors21, liu23}), but this could be remedied by models that allowed for higher disk density values than the standard value of $10^{15}$\,cm$^{-3}$ \citep{garcia18, tomsick18}. The disk density of $10^{15}$\,cm$^{-3}$ is a value appropriate for AGN accretion disks that spawned the field of reflection modeling, however, the accretion disks around stellar mass BHs and NSs are expected to have densities $\geq10^{20}$\,cm$^{-3}$ \citep{shakura73, Frank02}. \cite{garcia16, garcia18} provide more details about the effect of disk density on the reprocessed emission spectrum. Some flavors of \relxill have a variable disk density parameter, $\log(n_{e}/{\rm cm^{-3}})$, with an upper limit of 19 or 20 depending on the model. The development of \relxill models with densities $\log(n_{e}/{\rm cm^{-3}})>20$ is an active area of development (Ding et al. 2023, $submitted$), though there are some {\sc reflionx} grids currently available\footnote{\url{https://www.michaelparker.space/reflionx-models}} that can be blurred with relativistic blurring kernals \citep{tomsick18, Connors21, liu23}.

The ionization state of the accreting material can be constrained from modeling of the reflection spectrum. The ionization parameter, $\log(\xi/{\rm erg\ cm\ s^{-1}})$, is defined as $\xi=4\pi F_{x}/n_{e}$ based on the definition from \cite{tarter69} ($F_{x}$ is the illuminating flux and $n_{e}$ is the disk density), can vary to describe a nearly neutral (0) to highly ionized (4.7) medium. The ionization state of the medium changes the overall spectral shape of the reflection spectrum which also varies as a function of disk density (see Figure~3 of \citealt{Garcia22} for thermal illumination of disks of varying density). The ionization parameters of NS LMXBs are typically moderate depending on the spectral state ($\log(\xi)=2.3$-$4.0$: \citealt{cackett10, Ludlam17a, Ludlam19a})

Another explicit parameter of the reflection models is the reflection fraction. Note that the model creators refer to this parameter as $R_{\rm f}$ or $R_{\rm frac}$, but I will use $f_{\rm refl}$ to avoid confusion with parameters that use $R$ to refer to radius. Due to strong gravitational light-bending effects close to these accreting compact objects, $f_{\rm refl}$ is defined from the reference frame of the primary source as the ratio of emitted photons that will interact with the disk versus those that will be directly emitted to infinity \citep{dauser16}. Consequently, it is possible for this to have a value greater than unity if the location of the illumination source is closer to the accretion disk and deeper in the gravitational well where light-bending effects strengthen, since the majority of emitted photons would interact with the disk with relatively fewer emitted towards infinity. 

One special flavor of \relxill that has yet to be publicly released\footnote{The reflection model for UCXBs is available upon request to the author or J. A. Garc\'ia.} is that used for the subclass of LMXBs known as ultra-compact X-ray binaries (UCXBs). These systems have orbital periods of $<90$\,minutes that indicates that the donor star cannot be a main-sequence star but rather a degenerate stellar companion like a white dwarf. Since the donor has a significantly different chemical composition that is nearly devoid of hydrogen and helium while overabundant in carbon and oxygen \citep{nelemans03}, special models are necessary. {\xillverco} was created to be able to model these unique reflection spectra that have low energy relativistically broadened O lines with weak Fe~K lines \citep{madej11, madej14, Ludlam19b, Ludlam21, moutard23}. This model has some key differences in comparison to others, such as a variable abundance of carbon and oxygen, $A_{\rm \rm CO}$, rather than $A_{\rm Fe}$. The primary source input is a cut-off power law but the model also takes into account the local thermal radiation from below by the accretion disk itself, the ratio these components is defined by the `Frac' parameter. The rest of the parameters resemble those of the \relxill suite described above. More detailed information about the {\sc xillverCO} model can be found in \cite{madej14, Ludlam21}.

\begin{figure*}[t!]
\centering
\includegraphics[width=\textwidth,trim=2 3 0 2,clip]{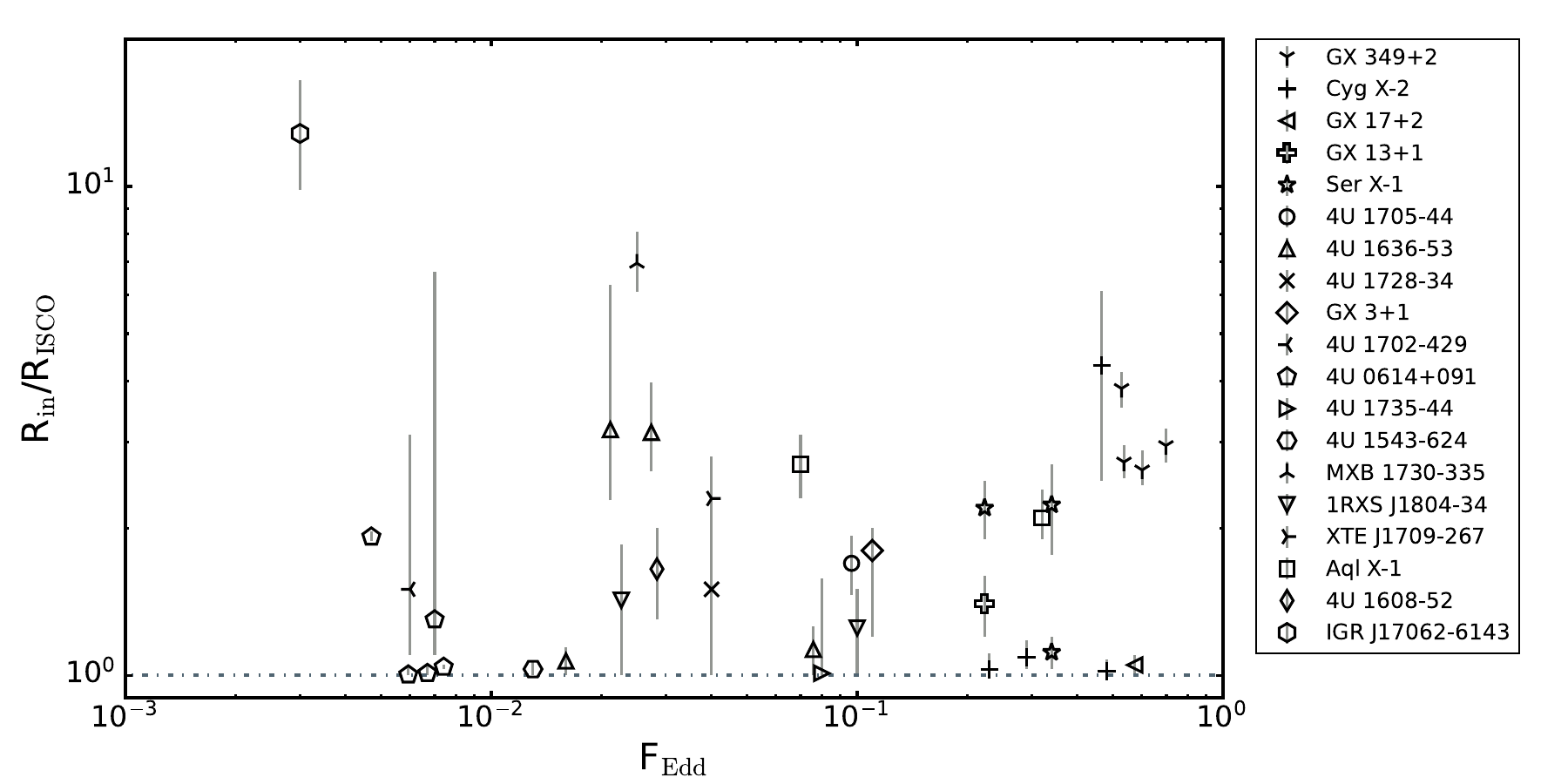}
\caption{Measured inner disk radius (\rin) in terms of \risco from reflection modeling versus the Eddington fraction (${\rm F_{Edd}}=L_{\rm 0.5-50\,keV}/L_{\rm Edd}$), which is a proxy for mass accretion rate (\mdot), for various NS LMXBs. Here, $L_{\rm Edd}$ is the empirical limit of $3.8\times10^{38}$ \lumcgs from \cite{kuulkers03}. The horizontal dot-dashed line indicated 1\,\risco. Figure adapted from \cite{moutard23}. These studies in particular were executed with \nustar observations and therefore assuredly unbiased by pile-up effects. These sources span several orders of magnitude in mass accretion rate, but no clear trend exists in terms of location of the inner edge of the accretion disk. References for the sources are: GX~349+2 \citep{Coughenour18}, Cyg~X-2 \citep{mondal18, Ludlam22},  GX~13+1 \citep{saavedra23}, Ser~X-1 \citep{miller13, matranga17, mondal20}, GX~17+2 \& 4U~1705-44 \citep{Ludlam17a},  4U~1636-53 \citep{Ludlam17a, wang17}, 4U~1728-34 \citep{sleator16}, GX~3+1 \& 4U~1702-429 \citep{Ludlam19a}, 4U~0614+091 \citep{Ludlam19a, moutard23}, 4U~1735-44 \citep{Ludlam20}, 4U~1543-624 \citep{Ludlam21}, MXB~1730-624 \citep{vanden17}, 1RXS~J1804-34 \citep{Ludlam16, Degenaar16}, XTE~J1709-267 \citep{Ludlam17b}, Aql~X-1 \citep{Ludlam17b}, 4U~1608-52 \citep{degenaar15}, and IGR~J17062-6143 \citep{vanden18}.}
\label{fig:fedd}
\end{figure*}

\subsection{Results of reflection modeling}

Reflection modeling has provided us with an additional method to infer the location of the inner disk radius based upon measuring the degree of relativistic effects impacting emission line profiles. As mentioned in \S~\ref{sec:intro}, measuring the location of the inner disk allows us to infer key properties of the NSs in these systems. There are a number of methods to determine the radii of NSs in order to determine the EOS of ultra-dense, cold matter (see \citealt{ozel16} for a review of different methodologies). Reflection spectroscopy offers an independent and complementary approach to the current best efforts that are underway (see Figure~5 in \citealt{Ludlam22} for a comparison to gravitational wave mergers and pulsar light curve modeling efforts).

An emission radius can also be inferred from the thermal radiation components (multi-color blackbody disk and single-temperature blackbody) in the spectrum. These rely on knowing the distance to the source, which can induce large uncertainties on the radius estimate in the absence of a precise distance measurement (e.g., see Table~5 in \citealt{Ludlam22}). Color correction factors are often applied when estimating the thermal emission radius to account for spectral hardening \citep{shimura95, kubota01}. For the disk blackbody component in particular, the inferred inner radius can be overestimated by a factor of $>2$ if not accounting for zero-torque inner boundary condition as expected for thin disk accretion in these systems \citep{zimmerman05}. Additionally, the inferred spherical emission radius of the single-temperature blackbody component can be  unphysically small for the emission to be  equivalent to the surface of the NS. If however the emission arises from a narrow banded region on the surface of the NS (i.e., the boundary layer/spreading layer; \citealt{inogamov99}) with a vertical height that is 5\%-10\% of the radial extent \citep{popham2001}, then this can increase the inferred radius to reasonable values \citep{Ludlam21}. The extent of the boundary layer region can also be estimated based on the mass accretion rate at the time of the observation by using Equation 25 of \cite{popham2001}. However, this does not account for the spin of the NS or viscous effects that can impact the radial extent of the boundary layer region. While all of these additional estimates aid in building up a coherent model of the accretion geometry in these systems, the advantage of reflection modeling to obtain the position of the inner disk radius relies heavily on proper modeling of the effects in the strong gravity regime near the NS.


The existing number of reflection studies in NS LMXBs utilizing data from \nustar (which are unbiased by pile-up effects) spans several orders of magnitude in mass accretion rate, \mdot. As a result, the search for a trend in the location of the inner disk radius as a function of \mdot has been investigated. As discussed in the introduction, the presence of reflection in NS LMXBs suggests a similar accretion geometry to BH systems. For BH XRBs, the accretion disk is thought to recede from the compact object at low \mdot (see \citealt{done07} for a schematic of the accretion geometry with spectral states/mass accretion rate). Figure~\ref{fig:fedd} shows the measured position of the inner disk radius as a function of Eddington Fraction, which is a proxy of \mdot since $L \propto$ \mdot. There is no clear correlation between position of the inner disk radius as a function of \mdot in these systems and this is likely due to the fact that NSs have a surface and an inherent magnetic field.

\begin{figure}[t!]
\centering
\includegraphics[width=0.48\textwidth,trim=10 10 0 10,clip]{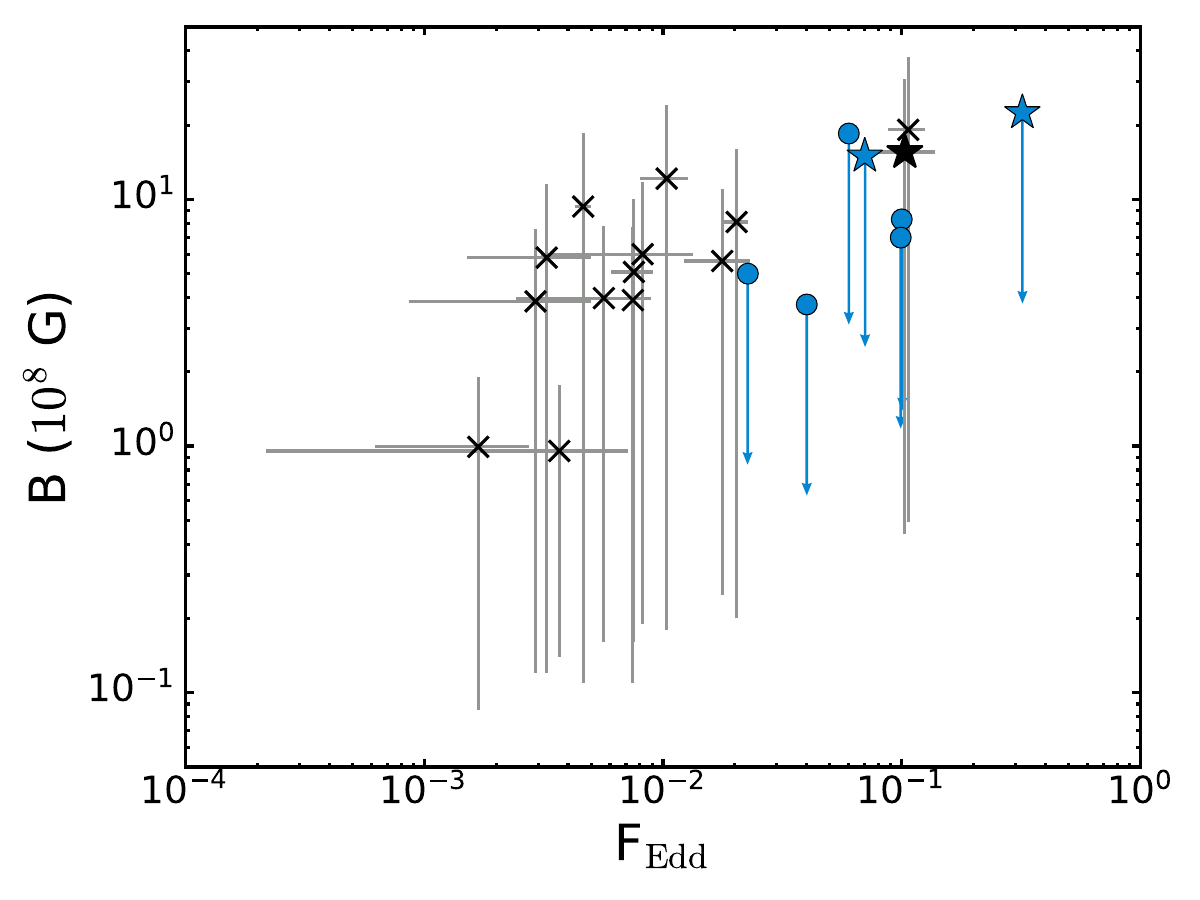}
\caption{Magnetic field strength of AMXPs (black crosses and star) in comparison to upper limits inferred from reflection modeling in several NS LMXBs (blue circles and stars). Figure recreated from \cite{Ludlam17c}. Of particular interest are the data points for Aquila X-1 (denoted as stars) which is an AMXP with an estimation from pulsations and reflection modeling. The agreement between the two methods for this source verifies that reflection modeling can be used to place limits on NS magnetic field strengths in the absence of discernible pulsations.}
\label{fig:mag}
\end{figure}

The strength of the magnetic field differs between sources (with $B$-fields on the order of $\sim10^{8-9}$\,G for LMXBs; \citealt{caballero12}), but can be strong enough to truncate the accretion disk before the ISCO. 
If the truncation radius inferred from reflection modeling is at the Alfv\'{e}n radius (where the energy density of the accretion flow equates to the energy density of the NS's magnetic field), then an upper limit on the strength of the magnetic field can be obtained \citep{ibragimov2009, cackett09}. This technique has been utilized to determine the magnetic field strength of NS LMXBs in the absence of pulsations, which could be shielded from our line of sight by several effects (see \citealt{lamb09} for more details). In particular, a study of the transiently accreting NS LMXB Aquila~X-1 (Aql~X-1) demonstrated the validity of using reflection modeling to obtain an upper limit on magnetic field strength \citep{Ludlam17c}. Figure~\ref{fig:mag} shows a comparison of estimates from reflection modeling in comparison to the magnetic field strengths from accreting millisecond pulsars (AMXPs) that show pulsations indicative of magnetospheric accretion that were reported in \cite{mukherjee15}. Aql X-1 is also an AMXP, so the agreement between the estimation of the $B$-field from these two methods lends credibility to using reflection modeling for gaining upper limits on magnetic field strength of NSs. 

However, the behavior of the inner disk radius with \mdot\ in individual systems needs to be considered to confirm if the magnetic field is responsible for truncation in the absence of detectable pulsations or if the extended boundary layer from the surface of the NS is responsible. If the magnetosphere is strong enough to impede the accretion flow, then the location of the Alfv\'{e}n radius changes as a function of mass accretion rate \citep{ibragimov2009}. If \mdot increases, then the energy density of the accretion flow increases and moves the disk closer to the NS. Conversely, if the boundary layer is responsible for truncation of the inner disk, in accordance with Eq.~25 in \cite{popham2001} the radial extent of the boundary layer grows as \mdot increases. In this case the inner disk radius would become more truncated. Hence, observing systems with truncated accretion disks over large changes in flux (and thereby \mdot) can be used to determine which mechanism is responsible. 

At low enough \mdot ($\rm F_{Edd}\lesssim10^{-2}$), standard thin disk accretion gives way to a radiatively inefficient accretion flow \citep{narayan94, esin97}. The exact \mdot value where this transition occurs is unclear, but monitoring transients as they return to quiescence after outburst provides an opportunity to capture observations in the lowest \mdot regime where this is expected to occur. Figure~\ref{fig:fedd} shows that some sources (e.g., 4U~0614+091) with $\rm F_{Edd}<0.01$ have a disk that extends down to \risco, so it must be possible to still have a standard thin disk down in this regime. Yet, \cite{vanden20} reported the disappearance of reflection features during the outburst decay of 4U~1608$-$52 that was observed at $\rm F_{Edd}\sim0.002$, which suggested that the accretion disk transitioned to the radiatively inefficient regime. Evidently, the accretion disk flow transitions somewhere between $0.002\lesssim \rm F_{Edd}\lesssim0.006$ and reflection modeling provides a means in narrowing this down further with more dedicated observations as sources decay from outburst. 


\section{Concluding Remarks} \label{sec:Conclusion}
Reflection model studies in NS LMXBs offer an independent method for determining upper limits on radii of NSs and their magnetic field strengths,  a probe into accretion when a surface is present (i.e., the presence of a boundary/spreading layer), the prospect to determine when standard thin disk accretion gives way to radiatively inefficient flows at low \mdot, and properties of the accreting material itself (e.g., density, composition, ionization state). Additional studies that combine X-ray polarization measurements with reflection studies can aid in furthering our understanding of both the illumination source (i.e., coronal orientation in line with jet or disk) and accretion geometry. These need to be modeled concurrently as the prescription of the energy spectrum does lend to the interpretation of the polarization measurements, hence reflection can provide an independent check in cases where the disk is inferred to be truncated from polarization properties (e.g., \citealt{Farinelli23}). 

Soon we will learn even more about these accreting systems through high energy resolution ($\lesssim5$\,eV) capabilities afforded by the successful launch of {\it X-Ray Imaging and Spectroscopy Mission} (\xrism; \citealt{xrism}), which is currently in the performance verification phase and set to begin community driven observations in 2024. The superior energy resolution of the microcalorimeter array onboard will be able to discern between minute differences in available reflection models and provide a more detailed view of the Fe line region to reveal structure due to density effects. In the late 2030s, the Advanced Telescope for High Energy Astrophysics (\athena; \citealt{athena}, recently re-scoped to {\it NewAthena}), is anticipated to become operational. The microcalorimeter array's energy resolution will be similar to \xrism, but the large collecting area and low background count rate of the Wide Field Imager (WFI; \citealt{wfi}) will enable targeting faintly accreting sources in less than half the exposure time needed for currently available facilities in order to understand accretion at low \mdot. 
To aid in how these missions are relevant to reflection modeling, \cite{Garcia22} provides simulated \xrism spectra showing the difference between two available reflection models and simulated O and Fe line emission profiles based on the anticipated capabilities with \athena. Reflection modeling is an advantageous technique to infer properties of accreting NSs and will undoubtedly enhance our understanding of these systems into the future.


\bibliography{references}

\end{document}